\documentclass[10pt]{article}
\usepackage{amsmath}
\usepackage{graphicx}

\newtheorem{protocol}{Protocol}

\title{A coercion-resistant protocol for conducting elections by telephone}
\date{5 November 2014}
\begin{document}
\maketitle 
\begin{abstract}
We present a protocol that allows voters to phone in their votes. Our protocol makes it expensive for a candidate and a voter to cooperate to prove to the candidate who the voter voted for. When the electoral pool is large enough, the cost to the candidate of manipulating sufficiently many votes to have an influence on the election results becomes prohibitively expensive. Hence, the protocol provides candidates no incentive to attempt inducement or coercion of voters, resulting in free and fair elections with the promise of cost savings and higher voter turnout over traditional elections. One major inadequacy with our suggested protocol is that we assume the existence of a trusted election authority to count the votes. 
\end{abstract}

\section{Introduction}
%
Being able to cast votes by telephone has some advantages over traditional voting in a polling booth: reduction in cost, higher voter turnout, and the possibility for true participative democracy are some that come to mind. Telephone voting also has some potential advantages over voting by internet, in terms of access and familiarity.

Secrecy of the vote is key to a free and fair election. With naive telephone voting protocols, this secrecy comes under threat if a candidate has a way of verifying who the voter voted for. Specifically, the candidate may recruit the voters' cooperation in such verification by coercion or inducement. This point appears to have been identified as early as 1994 by Benaloh and 	Tuinstra~\cite{benaloh1994receipt}. They coined the term ``receipt-freeness'' to describe the property that the voter should not be able to prove to another party who they voted for.

Since Benaloh and Tuinstra's work, a vast literature has accumulated around this problem. Techniques used include zero-knowledge proofs~\cite{jakobsson1996designated, baudron2001practical}, homomorphic encryption~\cite{hirt2000efficient,ETT:ETT4460080506}, blind signatures~\cite{okamoto1998receipt}, mix networks~\cite{sako1995receipt,juels2005coercion}, etc. An extensive survey on this literature has been done by Fouard, Duclos, and Lafourcade~\cite{fouard2007survey}. They conclude that ``it seems important not to lose sight of the fact that electronic voting schemes aim to be implemented and applied in ``real life'' with ``real people'' for elections.'' In this paper, we propose such a ``real life'' and pragmatic electronic voting scheme. 

We propose a protocol, and argue that with our protocol, it is expensive for a candidate to obtain from a voter proof of whom she voted for, even when the voter can be induced or coerced to cooperate. When the electoral pool is large enough, the cost to the candidate of manipulating sufficiently many votes to have an influence on the election results becomes prohibitively expensive. In practice, the candidate will have no incentive to attempt inducement or coercion of voters, even when voters are willing.

From the voter's end, our protocol is simple. Voters call a phone number. They are authenticated in a manner that requires nothing more from them than a password and replying to a voice prompt, and then cast their votes. Voters can call from any telephone device whatsoever, and no special features are needed. Our protocol promises reduction in cost of holding elections, higher voter turnout, and the possibility for the election commission to detect fraud through statistical analysis of voting patterns.

Our protocol is simple, natural, and concerned with practical implementability. We adopt a pragmatic, informal modeling that seeks to take into account real-world constraints on all actors. We argue the suitability of our protocol by an enumeration of scenarios. One major shortcoming with our suggested protocol is that we assume the existence of a trusted election authority. This is in contrast to traditional elections by ballot, which are arranged so that the voters and the candidates do not need to trust the election authority to believe the results. Till this serious limitation is addressed, the current protocol is not ready for adoption.

\section{Problem Statement}
We describe the \textbf{Phone Election Problem} of designing a protocol to allow voters to phone-in their votes. The actors in this protocol are:
\begin{enumerate}
\item The candidates: These are the parties contesting the elections. We assume candidates are evil: it is in the interest of each candidate to win the elections, and he will make \textbf{every effort} to do so. Candidates are very powerful, but not arbitrarily so.
\item The election authority: This is the body entrusted with the conduct of free and fair elections. It has limited resources with which to detect voter manipulation and electoral fraud. We assume that the election authority is trustworthy and above manipulation.
\item The voters: The voters communicate their choice of candidate to the election authority. We assume that each voter has a true choice, and a small incentive to communicate her true choice to the electoral authority. However, the introduction of a sufficient side incentive by one or more of the candidates may induce her to deviate from her true choice.
\end{enumerate}

For pedagogic reasons, we will focus on one candidate, and one voter. We will use masculine pronouns and adjectives for the candidate, feminine pronouns and adjectives for the voter, and impersonal pronouns and adjectives for the election authority. The aspects of this problem that interest us here are:

\begin{enumerate}
\item Voter manipulation: How do we design a protocol where the candidate has no incentive to manipulate the voter, either by coercion, or by inducement?
\item Authentication: How will the voter authenticate herself to the election authority? The election authority should be able to distinguish a voter from a proxy for the voter, {\em even when a proxy can enforce a voter's cooperation for short periods of time.}
\item Receipt-freeness: How can we prevent a candidate and a voter from cooperating in such a way that the candidate learns who the voter voted for?
\end{enumerate}

Towards these goals, we will describe a protocol (Protocol~\ref{profinal}). Our key idea is to make it so easy for a voter to change their vote that it becomes difficult for her to prove to somebody else who she voted for. The authentication is two-stage: a password and a voice-prompted speaker verification.

\section{No-Proxy Protocols}
The phone election problem has some novel features due to which many of the usual cryptographic primitives, like passwords and fingerprints, are inappropriate. We illustrate this issue with the example of a simple but flawed phone voting protocol. 

\begin{protocol}\label{pro1}[Phone Election 1]
The election authority has supplied the voter with a password. To vote, the voter calls a phone number, and authenticates herself using the password. Upon successful authentication, she gets to vote for the candidate of her choice.
\end{protocol}

The problem with this protocol is its vulnerability to voter manipulation. A candidate can manipulate the voter to disclose her password to him. Further, he clearly has incentive to do so -- once he has possession of the password, he can be a proxy for the voter, and cast her vote in his favour. 

Note that fingerprints, photographs, iris scans, voice recognition, etc., are ``secret keys'' --- static strings of bits that are to be communicated --- shared between the voter and the election authority. Hence, in principle, they could all be vulnerable to similar manipulation. Thus a key difficulty in the phone election problem appears to be the following: \textbf{how to authenticate an individual without making the authentication mechanism a commodity that can be traded.} We call this the \textbf{no proxy} problem, to capture the idea that only the voter herself should succeed at the authentication, not even a proxy appointed by the voter should succeed. 

Below we provide a voice-based protocol for the {\em no proxy} problem. The key idea is to require the generation of a string of bits on-the-fly. No one but the voter should be able to generate this string of bits on-the-fly. Hence, the proxy can succeed in the protocol only with the real-time assistance of the voter. This is not exactly what we need for the phone voting application. In some ways it is asking for too much, and in other ways it is not asking for enough. However, it will form a piece of our eventual protocol.

\begin{protocol}\label{pro2}[Voice-based no proxy]
As is usual for verification protocols, the protocol proceeds in two stages: the set-up stage where the secret to be shared is distributed, and the verification stage.
\begin{enumerate}
\item{Set-up:} The election authority collects a voice sample of the voter and assigns to her a unique secret voter id number. 
\item{Authenticate:} The voter or a proxy contacts the election authority, and makes a claim about her identity by entering her unique secret voter id number. A randomly-chosen paragraph from a large corpus is read out to her, and she is prompted to repeat it within a time limit. The election authority allows the vote to proceed if and only if 
	\begin{enumerate}
	\item The prompted paragraph was correctly repeated, and 
	\item The voice in which it was repeated matches the voice on file for that voter id number.
	\end{enumerate}
\end{enumerate}
\end{protocol}

We will assume that the candidate is not capable of manipulating the set-up step. One way to ensure this is to insist that for the set-up step, the voter go in person to the election authority, as in traditional voting, so that her identity can be verified, and voice sample collected. In principal, this step only needs to be done once in the lifetime of most voters. (However, note that voice aging is an issue; see Section~\ref{sec:vpsa} and Protocol~\ref{profinal} for a discussion and a proposed fix.)

It is important that the verification step in Protocol~\ref{pro2} be computationally feasible. We now argue that this is indeed the case. Note that the election authority has to perform two pieces of computation. 

The first is to verify that the prompted paragraph was correctly repeated. \textbf{This is a much easier problem than speech-to-text: all we are asking is whether the person on the phone is saying what she was prompted to say.} We are not required to produce an exact transcript of what was actually said, because we already know what the person ought to be saying. We only have to say ``yes'' or ``no'' depending on whether the text transcript of the voice is close enough to the original text. 

The second piece of computation is to match the voice with the voice on file, and return a ``yes'' or ``no,'' depending on a voice analysis. \textbf{Our protocol does not require voice recognition of one voice from among billions. All we are asking is whether the person on the phone is the person they claim to be.} This only requires, given two voice files, to say whether or not they belong to the same person. Even if two people, A and B, have the same voice, it doesn't matter. This is because, when A calls up to vote, she says that she is A, (or more accurately, enters her unique secret voter id number) and then proceeds to authentication. So her voice is not searched against B's voice sample, but only against her own voice sample on file. It is in this sense that our problem doesn't get harder as the number of people increase.

A potential problem arises only when B knows she has the same voice as A, so that she tries to steal A's vote. Even in this case, B will somehow have to get hold of A's unique secret voter id number. We discuss this issue in more detail when we come to Protocol~\ref{profinal}. Issues concerning the technical feasibility of voice-prompted speaker-verification, and robustness against voice forgery, 
will be examined in Section~\ref{sec:vpsa}.

There will be some voters (call them type II) who are vocally challenged, or develop vocal disabilities or alterations after their voice was sampled in the set-up stage. One can allow such voters to vote by a face-recognition based ``no proxy'' protocol, which requires video-enabled devices. The election commission may need to make some arrangements so that such voters have access to video-enabled devices. 

\begin{protocol}\label{proFace}[Face-based no proxy]
\begin{enumerate}
\item{Set-up:} The election authority collects photographs of the voter and assigns to her a unique secret voter id number. 
\item{Authenticate:} The voter or a proxy contacts the election authority on a video-enabled device, and makes a claim about her identity by entering her unique secret voter id number. A randomly-chosen paragraph from a large corpus is prompted to her. The election authority allows the vote to proceed if and only if (1) the prompted paragraph was correctly repeated as ascertained from lip movements, and (2) the face on camera matches the face on file for the voter.
\end{enumerate}
\end{protocol}

Alternatively, instead of prompting with words, the prompt could be for a sequence of facial gestures. For those voters (type III) for whom even this mode of voting is not possible due to more severe disabilities, traditional electoral booths or other forms of direct human verification may be arranged. Presumably voters of type II and III will be a small fraction of the population, so the cost of making such arrangements will be much less than the cost of traditional elections. In addition, many people who were previously unable to travel to electoral booths due to infirmities and disabilities will now be able to vote.

\section{Voice-prompted Speaker Authentication}\label{sec:vpsa}
In this section, we review the state of the art in voice-prompted speaker authentication.

The state of the art in forensic speaker recognition has been reviewed by Campbell et al.~\cite{campbell2009forensic}. In the next two paragraphs, we summarize the implications of their review to the technical feasibility of our proposal.

Recall that two metrics for the comparison of systems are ``false positives'' --- when the system accepts as a positive instance something that is not a positive instance --- and ``false negatives.'' Any system can be tuned to have a low percentage of false positives by allowing a high percentage of false negatives, and vice versa. So it is not meaningful to compare two systems based on only one of these two metrics. A non-trivial comparison can be obtained by looking at the percentage of errors at the ``equal error rate'' (EER) setting where the percentage of false positives and false negatives are equal. Campbell et al. report an EER of less than $2\%$, given a voice sample of around two and a half minutes. Possibly there is still room for improvement based on fine-tuning for the particular application we have in mind. We may be willing to accept higher false positives than false negatives, since people should not be denied their vote, even if the system becomes slightly easier to fool. 

Voice aging has been raised as a potential problem that may increase the EER. This claim has not been conclusively established, and appears to be a topic of ongoing research. If voice aging of the period of months seriously degrades system performance, that would be a concern for our application. However, voice aging of the period of years need not be a serious problem, provided elections are held frequently enough, allowing voice samples to be updated during phone voting. We are optimistic that existing systems can be made more robust to short-term voice aging.

Speaker recognition systems can be fooled, as demonstrated in \cite{bonastre2007artificial}. The idea is to apply a ``transfer function-based voice transformation'' by which an impostor's voice can be made to sound like a target speaker's voice. These systems seem to have an EER of around 50\%, so they still fail about half the time. In our application, there is an added requirement making voice forgery even more difficult. That is, the forged voice has to preserve intelligibility of what was said, otherwise the text-to-speech conversion will fail. Even if the systems of \cite{bonastre2007artificial} can meet this additional requirement, there are two ways for the election authority to attempt to thwart the threat of voice forgery, and decrease the candidate's confidence that a particular voice forgery system is working.
\begin{enumerate}
\item \textbf{Honeytrap:} If the voice forgery system has a distinctive signature, then that may show up in a statistical analysis of all the votes. The election authority may ``pollute'' the market for voice forgery systems by putting out voice forgery systems with such signatures, so that unwary candidates end up using such signed systems to manipulate elections, and get detected.
\item \textbf{Secret features:} A voice forgery system appears to require some knowledge about the targeted speaker recognition system. The election authority could keep the details of the speaker recognition system secret. In addition, they could keep more than one speaker recognition system, randomly employing any one of these for a particular vote authentication. It is important that the candidate not be able to get any feedback about whether a particular attempt at authentication is successful, to prevent fine-tuning. Therefore, the system should appear to behave exactly the same with an authentic user and someone who has failed the authentication, in appearing to allow them both to vote. At the same time, we would like voters to be able to keep track of whether their vote was successfully cast, if they wish to do such tracking. We will return to this issue when we disuss receipts in Protocol~\ref{profinal}
\end{enumerate}

\section{Phone Election Protocol}
The voice-based {\em no proxy} protocol  may not be directly suitable for use for the phone election problem. This is because if only the first vote is to be counted, a candidate can try to make sure the voter votes for him before she has a chance to vote for anyone else. We will circumvent this problem by allowing a voter to vote multiple times, once across every phone session, and only count her last valid vote. This is the main idea of our paper.

The candidate can now try to make sure that the voter's last vote is cast in his favor. To ensure this, he can arrange to have the voter vote for him just before voting closes, and then for the rest of the time, make sure that the voter does not have access to a phone, for example, by keeping her locked in a room. Another possible manipulation is to get the voter to consume a ``poison'' just after she votes for the candidate in his presence, so that she acquires a voice handicap and is unable to change her vote. We will refer to strategies of this nature as ``sequester manipulations.''

Note however, that the candidate needs to be able to manipulate the votes of a large number of voters to make an impact on the result of the elections. Let us be pessimistic, and assume that he can manage to arrange the simultaneous sequester manipulation of a large number of voters. The election authority can make the cost of his playing the ``sequester manipulation'' strategy very high, by making the stopping time of the election a random variable. As a result, to successfully manipulate an election result, the candidate has to play the sequester manipulation strategy for a long time, as well as on a large number of voters. Thus a random stopping time has two disadvantages for the candidate:
\begin{enumerate}
\item Sequester manipulation becomes more expensive.
\item Harder to avoid detection of the manipulation by the election authority.
\end{enumerate}

Here is what me mean by a random stopping time. The election authority will only announce that voting will be open till a certain time epoch. After that time epoch, the stopping time of the elections will be distributed according to some uniform random variable with a sufficiently high mean, say a week. The election authority will sample this random variable, and keep the sample secret. The extended period of time that the candidate will be required to play the sequester manipulation strategy will also increase the opportunity for the election authority to become aware of such manipulation, and take appropriate action.

To make this concrete, consider a time epoch $\tau = 200$ hours. The election commission collects a sample $x$ of a uniformly-distributed random variable $X$ taking values in the interval $[0,1]$. The secret stopping time is chosen as $\tau + 200 * x$. Most voters who are not under coercion will simply ignore the stopping time and vote before time $\tau$ to be assured that they can cast their vote. So the secrecy of the stopping time does not hinder anyone from voting. The election abruptly stops at the stopping time, which has an expected value of $200 + 100$ hours or roughly $1.7$ weeks, but can take any value between $200$ hours to $400$ hours.

The final protocol is as follows.

\begin{protocol}\label{profinal}[Final Protocol]
\begin{enumerate}
\item{Set-up:} The election authority collects a voice sample of the voter, and assigns to every voter a unique secret voter id number.
\item{Stopping time:} The election authority announces elections, and a time epoch $\tau$  --- say $\tau=200$ hours for concreteness --- till which time voting will remain open. Given a sample $x$ from a uniformly-distributed random variable $X$ taking values in the interval $[0,1]$, the stopping time is chosen as $\tau + 200*x$. The sample $x$ is kept secret with the election authority.
\item{Authenticate:} While the stopping time has not been reached, the voter or a proxy contacts the election authority, and makes a claim about her identity. Her unique secret voter id number is verified. Then a randomly-chosen paragraph from a large corpus is read out to her, and she is prompted to repeat it within a time limit. The election authority authenticates the voter if and only if
	\begin{enumerate}
	\item The password was correct.
	\item The prompted paragraph was correctly repeated, and 
	\item The voice in which it was repeated matches the voice on file for the voter.
	\end{enumerate}
\item{Vote:} An authenticated voter is allowed to vote, and the voice sample collected is appended to the voice sample on file. A non-authenticated voter is also allowed to vote, without any indication being given to the voter that the authentication has failed, but this non-valid vote is immediately discarded.
\item{Receipt:} The voter is asked if she would like a receipt. If the voter is an authenticated voter, then the choice is remembered. The voter can either say ``yes, I want a receipt'' or ``no, I don't want a receipt'' or she can skip these options. If she skips, the setting that was made on the previous call (or the default setting of ``no receipt'' if this is the first call) will be preserved. No receipts are sent out at this point, only the choice is remembered.
\item{High scrutiny:} An authenticated voter is allowed to set her voter id to ``high scrutiny'' in case she believes that her id is particularly vulnerable to an attack, say because she has been coerced to part with her unique secret voter id number.
\item {Last vote counts:} While the stopping time has not been reached, a voter may vote multiple times. Each vote requires a fresh authentication. Only the voter's last authentic vote before the stopping time is taken into account when tallying results.
\item {Post-election:} After voting has ended, if, and only if, a voter's receipt setting is set to ``yes, I want a receipt,'' she receives exactly one receipt from the election authority. The receipt says that she voted, and nothing more. 
	\begin{enumerate}
	\item Even if an authentic voter voted many times, and asked for a receipt each time, she never receives more than one receipt. 
	\item If she asked to not receive a receipt the last time she was successfully authenticated and voted, she does not receive any receipt, irrespective of whether or not she asked for a receipt on previous votes. 
	\item If no authentic vote was cast from her id, she does not receive any receipt.
	\end{enumerate}
\item {Fraud detection:} The election commission will conduct statistical analysis on the vote patterns to see if any manipulation can be discerned. Some simple things that can be tried include clustering of voices, to see if many voices can be attributed to the same voice forger, geographical and phone number based clustering, to see if unreasonably many votes were cast from the same or nearby phone numbers, etc. Special care will be taken with votes from numbers that have been marked as ``high scrutiny.''
\end{enumerate}
\end{protocol}

The election authority will also submit to higher scrutiny devices from which a large number of votes have been cast, or are likely to be cast, possibly sending their representatives to directly supervise such devices.

\subsection{Analysis of Protocol~\ref{profinal}}
If a voter is not willing to cooperate with the candidate, then he can't manipulate her vote, because she will keep her password secret, so stealing her vote is as hard as stealing money from her bank account. If the voter wants to know if her vote was successfully cast, she can ask for a receipt. If she receives it, she knows that her vote was cast. If she is confident in the secretness of her unique secret voter id, then that is the end of the matter. 

Imagine that a candidate tries to ``buy'' a vote by rewarding a voter for voting for him. It is very hard for the voter to prove to the candidate that her vote is indeed cast in his favour. This is because it is very easy for the voter to change her vote; all she needs is access to a telephone. Therefore, how does the candidate know that the voter will not defect from their agreement? Since the candidate has no way of verifying who the voter voted for, he has no incentive to attempt manipulation, even if the voter herself offers to cooperate with him for a small incentive.

We now examine other attacks the candidate can perform.

\subsubsection{Voice forgery attack}
If the candidate could forge a voter's voice, he will still be able to manipulate her vote.  Let us pessimistically assume that the candidate is in possession of a perfect and undetectable voice forger. This is indeed a pessimistic assumption because even the voice forgers of \cite{bonastre2007artificial} only succeed 50\% of the time, and they have some knowledge about the speaker recognition system.

If the voter is willing to cooperate, the candidate pays the voter a small amount to acquire her voice sample and password. The voter could bluff the candidate, either by giving him a false, forged password, or by using a voice sufficiently different from her normal voice. The candidate gets no feedback from the telephone system, so he has no way of knowing if his perfect voice forger is correctly getting authenticated. As a result, he is vulnerable to such bluffing. He is also unable to ascertain if his voice forger is working properly, or to improve it.

Now he needs to keep voting again and again on behalf of the voter to make sure the vote stays, especially in the period after the time epoch $\tau$ and before voting closes. Meanwhile, the voter knows the candidate is repeatedly voting to manipulate her vote. So she can vote many times to try to recover her vote. She can put her voter id number on high scrutiny. So even to manipulate one vote successfully is quite expensive for the candidate. It gets worse, because the candidate has to successfully manipulate a large number of votes to make a difference to the outcome of the election. All these votes can not be cast from one phone number, else the election authority may get suspicious. If the perfect voice forger has a signature that is common across all its forgeries, even that can be detected by the election authority.

One may ask, if a candidate is indeed able to obtain a perfect and undetectable voice forger, what is the point of having a voice-prompted speaker authentication step in our protocol at all? Why not simply have a password, and repeated voting, with a ``last vote counts'' rule? One answer is that at least with current technology, perfect voice forgers are semi-automated systems, not fully automated. That is, the candidate would listen to the voice prompt, and repeat it, and the voice forgery system would transform what the candidate said to sound like the voter said it. This makes it necessary to have a human in the loop, doing the voting and revoting for each voter, and limits the power of the candidate. Otherwise, if the whole process could be automated, one could imagine a fully-automated, distributed attack that would be harder to detect, and be less expensive. 

Thus, part of the job being done by the voice-prompted speaker authentication is like that of a ``captcha''~\cite{von2003captcha}: a Turing test to distinguish humans from computers, and slow down credential forgery. One can make this part more secure by adding some garbling sound signals to the prompt, making it harder for a computer to parse the prompt. Since humans can perform error-correction based on knowledge of the semantic and pragmatic content of the sentence, they may be able to tolerate higher levels of error than computers. One hopes that by the time perfect voice forgers become completely automated and run in real-time, video on phones will have become ubiquitous so that one can move over completely to video-based protocols instead of voice-based ones.

\subsubsection{Denial of vote attack}
The candidate can try to coerce a voter to not vote. But again, the voter may vote and ask to not be sent a receipt, and the candidate has no way of knowing if the voter voted or not. The candidate may try jamming phone signals, or sound pollution, or a sequester manipulation. All these have to be carried out on a scale that is large, while at the same time remaining discreet so as not to be detected by the election authority. Note in particular that if no votes are being cast from a large geographical area, that would be detected by the election commission.

In traditional elections, people who want to vote may be prevented from reaching the polling booth. This problem is reduced in our scheme, because voting is no longer done in a geographically localized manner.

\subsubsection{Bluff manipulation}
We describe another possible manipulation that we call the ``bluff manipulation.'' In this strategy, the candidate falsely claims to have knowledge over who voted for and against him. After the election, he randomly picks a voter from the population, and punishes the voter irrespective of who the voter actually voted for. Because the vote was secret, only that voter can possibly know that the candidate was bluffing, and the candidate may still be able to intimidate the rest of the population into voting for him. This sort of manipulation is a weakness shared by traditional election methods too, so our protocol doesn't seem to be more vulnerable to such manipulation than the traditional paper ballots.

\subsubsection{Channel attack}
We have assumed that the telephonic network over which this exchange is taking
place can be trusted. One can imagine an attack where the network is
compromised, or the telephone instrument used to cast votes is compromised.
We have not considered such attacks. We have implicitly assumed that the election authority takes 
necessary steps to secure these networks, and educate voters to secure their devices. If voters can perform certain cryptographic operations, then it should be possible to secure the protocol against channel attacks also. Below we briefly describe ideas for countering such attacks.

To prevent eavesdroppers from learning her vote, a voter may send an encrypted version of the vote. It is important that encrypted versions of the votes of different voters voting for the same candidate should not be identifiable by an eavesdropper as votes in favor of the same candidate. 

The eavesdropper may not care to learn the vote, but may care only to change the vote, by intercepting the sent message, and sending a different message to the election commission. If an eavesdropper is manipulating a large fraction of the messages, then the election commission can  detect this by having their own agents cast votes over the channel. If what they receive is maliciously different from what they except to receive, they should suspect the presence of an eavesdropper.

\subsubsection{Small population attack}
When the population size is small, the election authority can increase scrutiny because the number of voters is smaller. Nevertheless, our scheme may be unsuitable because candidates may be able to afford to manipulate sufficient votes to make a difference to the election results. But then, even in traditional elections, when the population size is small, the result of the election is informative about who a voter voted for, and so votes are only partially secret. 

The ease of voting in our scheme may translate into higher voter turn-outs. When more people vote, the election is inherently more expensive to manipulate, because changing the result requires changing more votes.

\subsubsection{Stopping time manipulation}
What if the election authority has not really sampled the random variable $X$ before the election starts, but instead closes the election abruptly when its favored candidate is in the lead? This can be prevented by a commitment scheme, along the lines suggested below.
\begin{enumerate}
\item The election authority creates a secret original document $O$ that has embedded in it, among other sufficiently random strings, the time of stopping of the elections.
\item The election authority uploads to its website before the start of elections an encrypted document $D$, produced by encrypting $O$ with the election authority's public key.  
\item After the elections are closed, the election authority publishes the original document $O$ on its website. Anyone can encrypt using the public key to verify that the encrypted version $E(O)$ of the original document $O$ does indeed correspond to the document D that the election authority made public.
\end{enumerate}

Thus, the election authority can not change the stopping time arbitrarily once elections have started, even though the document $O$ remains secret. It is important that along with the random stopping time, $O$ also have other random strings, because otherwise someone can try to look for encryptions of all possible stopping times, and compare with the encrypted document $D$, to recover the original stopping time.

\subsubsection{False counting}
Perhaps part of the faith that people have in the security of traditional elections has to do with the fact that there is no single point of failure. Many people are always around the ballot boxes, and the counting of votes happens in a room full of people where it is very difficult to fudge results. One concern might be whether we have lost such robustness by relying so highly on a trustworthy election authority. This is not the main issue discussed in this paper, and is in fact a separate issue that deserve a detailed consideration, along with several other aspects like the psychological and political acceptability of our protocol, and the social implications. Below we indicate one way towards addressing this concern.

The idea is to have multiple election authorities. Before elections begin, the electorate is randomly partitioned in equal numbers among the different authorities. The splitting has to be random enough, and the population size large enough, that we believe all the authorities are sampling from the same distribution of voters. When a voter phones in, after she enters her electoral id number to establish who she is claiming to be, she is routed to the appropriate election authority. Then each authority separately counts its votes, and uses a commitment scheme to publish its count. Finally after all the encrypted counts have been published, all the counts are revealed. So long as the majority of election authorities have not colluded together to fudge the results, such a scheme can potentially be made robust to false counting by a small number of election authorities, because the counts of the false-counters will be far away from the counts of the others.

\subsubsection{Voting confirmation}
Voters may want confirmation that their vote was counted. As we have already noted, if the system gives immediate feedback about whether a vote is accepted, this can be used against it by voice forgery systems as a feedback loop to figure out how to break the system. Our system provides a weaker feedback. The feedback is only available after the elections are over. It does not give information on each attempt at voting, but only on the last successfully authenticated vote cast by the voter. We expect most voters will not be concerned about coercion. They will only vote once, and want to know that the vote they cast was valid. So they will ask for a receipt. In the normal use case, the receipt will be received by them after the elections, confirming that their vote was cast successfully. If they asked for a receipt, and do not receive it, then they know something went wrong. It is too late to fix things in the last election, but they are at least aware there was a problem, and can try to fix things before the next election. If many people have a problem, they can come together, and voice a protest. At the same time, if a candidate is coercing a voter to not vote, she can vote and ask to not receive a receipt, so her having voted remains a secret.

Would such a system be perceived as fair, if people learn about whether their vote was validly cast only after the elections are over, and it is too late to do anything about it? Consider that technical issues happen all the time during polling. There are many citizens whose names do not appear on the electoral rolls, or who do not have voter ID cards. They go to the polling booth and are not allowed to vote. Sometimes votes get annulled because of technical issues like ``dimpled chads'' in the state of Florida in the 2000 United States presidential elections. Voters denied their vote do express their displeasure, but there are few calls to overthrow the entire system. There seems to be an acceptance that holding elections is a complex activity prone to errors. So long as those errors are not a priori biased in favor of one outcome, people seem to take such errors in their stride. Thus our protocol aspires to a version of fairness that is comparable to ``envy-freeness''~\cite{1995}: it is not a priori unfair to any one of the candidates.

\section{Conclusion}
The problem of implementing electronic elections is one of practical significance. We have begun a conversation towards pragmatic protocols for this problem, taking into account the messy real-world details necessary to actually make such schemes work. One way forward is to build prototype systems, and test and prove them in suitable settings. Besides the technological aspects, there are also psychological, sociological, and political aspects of such schemes that require careful study that is beyond the scope of this paper. 

\bibliographystyle{amsplain}
\bibliography{references}
\end{document}